\documentclass[journal,twoside,web]{ieeecolor}
\usepackage{cite}
\usepackage{tmi}
\usepackage{amsmath,amssymb,amsfonts}
\usepackage{algorithmic}
\usepackage{graphicx}
\usepackage{textcomp}
\def\BibTeX{{\rm B\kern-.05em{\sc i\kern-.025em b}\kern-.08em
    T\kern-.1667em\lower.7ex\hbox{E}\kern-.125emX}}
\begin{document}
\title{Sparse-view Signal-domain Photoacoustic Tomography Reconstruction Method Based on Neural Representation}
\author{Bowei Yao, Yi Zeng, Haizhao Dai, Qing Wu, Youshen Xiao, Fei Gao, YUyao Zhang, Jingyi Yu, and Xiran Cai
\thanks{Yi Zeng, School of Information Science and Technology, ShanghaiTech University, Shanghai, China}
\thanks{Haizhao Dai, School of Information Science and Technology, ShanghaiTech University, Shanghai, China}
\thanks{Qing Wu, School of Information Science and Technology, ShanghaiTech University, Shanghai, China}
\thanks{Youshen Xiao, School of Information Science and Technology, ShanghaiTech University, Shanghai, China}
\thanks{Fei Gao, School of Information Science and Technology, ShanghaiTech University, Shanghai, China}
\thanks{Yuyao Zhang, School of Information Science and Technology, ShanghaiTech University, Shanghai, China}
\thanks{Jingyi Yu, School of Information Science and Technology, ShanghaiTech University, Shanghai, China}
\thanks{Xiran Cai, School of Information Science and Technology, ShanghaiTech University, Shanghai, China}
}

\maketitle

\begin{abstract}
Photoacoustic tomography is a hybrid biomedical technology, which combines the advantages of acoustic and optical imaging. However, for the conventional image reconstruction method, the image quality is affected obviously by artifacts under the condition of sparse sampling. in this paper, a novel model-based sparse reconstruction method via implicit neural representation was proposed for improving the image quality reconstructed from sparse data.  Specially, the initial acoustic pressure distribution was modeled as a continuous function of spatial coordinates, and parameterized by a multi-layer perceptron. The weights of multi-layer perceptron were determined by training the network in self-supervised manner. And the total variation regularization term was used to offer the prior knowledge. We compared our result with some ablation studies, and the results show that out method outperforms existing methods on simulation and experimental data. Under the sparse sampling condition, our method can suppress the artifacts and avoid the ill-posed problem effectively, which reconstruct images with higher signal-to-noise ratio and contrast-to-noise ratio than traditional methods. The high-quality results for sparse data make the proposed method hold the potential for further decreasing the hardware cost of photoacoustic tomography system.
\end{abstract}

\begin{IEEEkeywords}
Photoacoustic tomography, Image reconstruction, Implicit neural representation, Multi-layer perceptron.
\end{IEEEkeywords}

\section{Introduction}
\label{sec:introduction}
\IEEEPARstart{P}{hotoacoustic} compute tomography (PACT) is a medical imaging technique which combines the contrast of optical imaging and the imaging depth of ultrasound imaging\cite{xu2006photoacoustic,wang2009multiscale,jathoul2015deep}. This technique uses short-pulse (nanosecond duration) laser to illuminate biological tissue and employees ultrasonic transducer to acquire ultrasonic wave generated by instantaneous thermal variation from light-absorbing structures, namely photoacoustic (PA) signals.
PACT enables functional and molecular imaging capabilities, which is widely applied in preclinical studies in the fields of tissue imaging, cancer detection and neuroscience research\cite{b1,b2,b3,mehrmohammadi2013photoacoustic,wu2022system,tian2015imaging, tian2016plasmonic}.
For in-vivo imaging, PACT has been used to acquire cross-section images and whole-body dynamics of small-animal \cite{b4,b5,zhu2024longitudinal,yeh2017dry}, periphery blood vessel structure of human limbs such as lower leg, finger and breast \cite{b6,b7,b8}, as well as human cerebral vasculature\cite{na2022massively}.

High quality PACT imaging usually requires a dense spatial sampling around the object to satisfy the Nyquist sampling theorem and to avoid spatial aliasing\cite{Hu2020Spatio}.
Thus, a high-channel count acquisition system and a transducer array of large number of transducer elements are usually deployed for imaging, resulting increased cost and complexity of the PACT system.
To eliminate the artifacts in the images caused by sparse data, such as the streak-type artifact, various approaches have been adopted.
Sacrificing spatial resolution of the image, the PA signals may be low-pass filtered so that the required spatial sampling frequency may be lowered as the wavelength used by the universal back-projection (UBP) method\cite{b12} for image reconstruction shall be increased\cite{Hu2020Spatio,b5}.
Another approach to improve the image reconstruction quality in PACT is using the model-based (MB) inversion schemes\cite{b10}, i.e., minimizing the loss function between the measured PA signals and the theoretical ones predicted by a certain PA forward model. 
Applying proper regularization terms and optimizing method, MB methods yield better image quality than the UBP method, even for a low number of sensors\cite{paltauf2002iterative}.
The trade-off is, however, the associated high computational cost. And the sparsity of sampling will yield the ill-pose problem, which impacts the solution accuracy of inverse problem. For relieving ill-posed problem, proper regularization terms and parameters are selected as constraint to accelerate the convergence of solution\cite{dong2015algorithm}.   
Deep learning based methods have been also introduced to improve image reconstruction quality from sparse data, using the networks based on U-net structure\cite{tong2020domain ,guan2020limited, davoudi2019deep} or attention-steered network\cite{guo2022net} in a supervised manner.
Thus, the training process requires a large dataset which is generally difficult to obtain in PACT and the generalization capability of the method heavily depends on the quality of the dataset.
Integrating a diffusion model to the inversion of MB method, the optimization problem for image reconstruction with sparse data can be better constrained\cite{song2023sparse}.
However, the diffusion model essentially learns the prior information of the data distribution to better constrain data consistency.
The method may only applicable to the learnt data distribution while not to other scenarios. For learning the data distribution, the training process of diffusion model takes for hours, which reducing the timeliness of image reconstruction. 

A common feature in all the currently adopted strategies to overcome the ill-pose problem of the inverse problem in PACT, particularly under sparse-view, is that the inversion is formulated in a discrete framework, i.e., the images to be optimized are represented as discrete matrix.
Thus, the reconstruction is prone to discretization errors at low resolution.
At high-resolution, the increased number of unknowns imposes high time and space complexity and makes the inverse problem more ill-posed.

In recent years, a new paradigm to formulate the inverse problem has emerged in the computer vision and graphics communities\cite{b17}. Based on optimizing a multi-layer perceptron (MLP), implicit neural representation (INR) models and represents 3-D scenes from a sparse set of 2-D views in a self-supervised fashion. Benefiting from the image continuity prior imposed by the neural network, INR has achieved superior performance for various computer vision tasks\cite{b17,zhang2020nerf++,muller2022instant}. Specially, by adding prior information provided by the regularization term, INR has achieved novel view synthesis from sparse-view data\cite{wang2023sparse, niemeyer2022reg}.   

Initiated by the computer vision and graphics communities, INR has been widely applied in other image reconstruction tasks. In computed tomography (CT) image reconstruction, INR methods were used to recover high-quality artifact-free images from a sparse-view sinogram data\cite{shen2022nerp,sun2021coil,wu2023self}. In non-line-of-sight (NLOS) imaging, INR method achieved state-of-the-art reconstruction performance under both confocal and non-confocal settings\cite{shen2021non}.


Enlightened by the aformentioned work, in this work, we proposed a framework based on neural representation(NR) for PACT image reconstruction with sparse data.
Specifically, the initial heat distribution is represented implicitly with a neural network whose parameters are determined by the radio frequency (RF) PA signals after training.
The training process essentially involves minimizing the errors between the measured PA signals and the signals predicted by a forward model relating the neural representation of the initial heat distribution to PA signals in self-supervised manner.
After the training, the PA images can then be mapped by feeding the coordinates to the network.
In the following, we present, evaluate and validate the proposed framework with the comparison of the reference image reconstruction methods (UBP and MB) using both simulation and experimental data for different spatial sampling conditions.
The results show that the proposed method performed best in terms of image fidelity and artifacts suppression than the reference methods for the same sparse view sampling conditions.

\section{Materials and Methods}
\subsection{Photoacoustic Wave Equation}
In photoacoustic tomography, nanosecond laser pulses satisfying the condition of thermal confinement and of neglecting heat conductance are used. In this way, the thermal expansion of the irradiated region is not affected by its neighboring regions\cite{xu2006photoacoustic}. The temporal profile of the light source can be then appreciated as a Dirac delta function and the pressure of the generated ultrasonic waves in a homogeneous medium is given by
\begin{equation}
\frac{\partial^{2} p(\boldsymbol{r},t)}{\partial t^{2}} - c^{2} \nabla ^{2} p(\boldsymbol{r},t) = \Gamma H(\boldsymbol{r}) \frac{\partial \delta(t)}{\partial t}\label{eq1}
\end{equation}
where $c$ is the speed of sound (SoS) of the medium, $\Gamma$ is the Grueneisen parameter and $H(\boldsymbol{r})=\mu_{a}(\boldsymbol{r}) U(\boldsymbol{r})$ is the amount of energy absorbed per unit volume  at position $\boldsymbol{r}=(x,y)$ with $\mu_{a}(\boldsymbol{r})$ the optical absorption coefficient and $U(\boldsymbol{r})$ the light fluence, and $p(\boldsymbol{r},t)$ represents the pressure field at $\boldsymbol{r}$ and time $t$. Considering the initial conditions
\begin{equation}
    p(\boldsymbol{r},t)|_{t=0} =  \Gamma H(\boldsymbol{r})\label{eq2}
\end{equation}
and
\begin{equation}
   \frac{\partial p(\boldsymbol{r},t)}{\partial t} \bigg| _{t=0} = 0 \label{eq3}
\end{equation}
 Eq.(\ref{eq1}) can be solved by Green's function method. $p(\boldsymbol{r},t)$ can be expressed as:
\begin{equation}
    p(\boldsymbol{r},t) = \frac{1}{4\pi c}\frac{\partial }{\partial t}\int_{S(t)} \frac{p(\boldsymbol{r'})}{|\boldsymbol{r}-\boldsymbol{r'}|} d S(t)\label{eq4}
\end{equation}
where $S(t)$ represents a spherical wavefront originated from $r'$ with $|\boldsymbol{r}-\boldsymbol{r'}| = ct$. 
\begin{figure*}
\centering
\includegraphics[width=\textwidth]{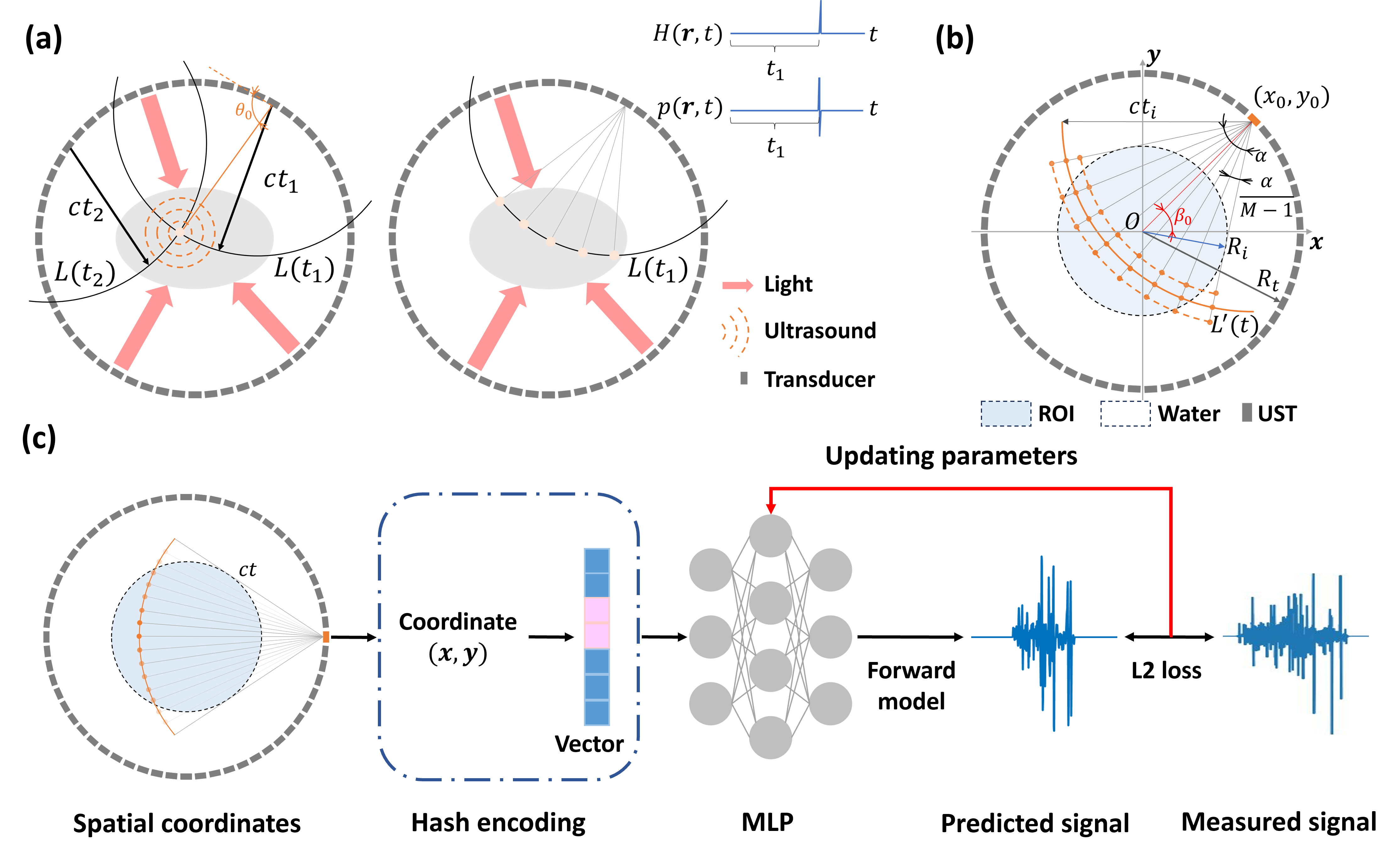}
\caption{Overview of neural representation for PACT. (a) Schematic demonstration of propagation of sound wave excited by laser. Under the circumstance of uniform speed of sound, the location of sound source (flesh color point) in body (grey ellipse) can be calculated form transducer element position and time-of-flight. (b) Temporal and spatial discretization of PA forward model. Points on orange solid line represent the sampled coordinates, and points on dash line represent the calculated coordinates.(c) The flowchart of PACT-NeRF proposed in this work. Plane coordinates inside ROI are encoded and put into network, which map the coordinates into heat. Subsequently, forward model maps the heat into ultrasound signal, which is calculated loss from real signal and the loss is back propagated to optimize network. ROI, region of interst; UST, ultrasound transducer.}
\label{Fig1}
\end{figure*}
\subsection{Universal Back-projection Method}
In 2D PACT with a ring transducer array of radius $R$ (Fig.~\ref{Fig1}), image reconstruction can be treated as solving the inverse problem of Eq.~(\ref{eq4}), i.e. reconstruct the initial pressure $p(\boldsymbol{r'})$from the RF data $p(\boldsymbol{r},t)$ collected by the ring array.
The universal back-projection (UBP)\cite{b12} method to solve the inverse problem is expressed as:
\begin{equation}
    p(\boldsymbol{r'})= \frac{1}{\Omega_{0}}\boldsymbol{\int}_{S'(t)}\big[2p(\boldsymbol{r},t) - \frac{2t\partial p(\boldsymbol{r},t)}{\partial t}\big]\frac{\cos{\theta_{0}}}{|\boldsymbol{r}-\boldsymbol{r'}|^{2}}dS'(t)
    \label{eq5}
\end{equation}
where $\theta_{0}$ is the angle between the vector pointing to the source position $r'$ and transducer surface $S'(t)$ (Fig.~\ref{Fig1}(a)). ${\Omega_{0}}$ is a solid angle of the $S'(t)$ with respect to the wave source. For the planar geometry, ${\Omega_{0}} = 2\pi$, and ${\Omega_{0}} = 4\pi$ for ring array with the cylindrical geometry used in this work. 

.

\subsection{Model-based Method}
Under the condition of full and dense view sampling, Eq.~(\ref{eq5}) can be approximately implemented by adding the projection data simply. However, the artifact occur for the condition of limited-view or sparse-view.
For enhancing the image quality under the non-ideal conditions, compared to the UBP method, model-based (MB) solves inverse problem for iteration. And regularization term is applied to improve the reconstructed image quality \cite{b10}.
For numerical evaluation of the forward model of PA wave propagation, the derivative in Eq.~(\ref{eq4}) is approximated as: 
\begin{equation}
    p(\boldsymbol{r},t) = \frac{I(t+\Delta t) - I(t-\Delta t)}{2\Delta t}\label{eq6} 
\end{equation}
where 
\begin{equation}
    I(t) =  \int _{L'(t)} \frac{H(\boldsymbol{r'})}{|\boldsymbol{r-r'}|}dL'(t)\label{eq7}
\end{equation}
which is discretized as
\begin{equation}
    I(t) \approx  \frac{1}{2}\sum _{l=1} ^{M-1} \Big [\frac{H(\boldsymbol{r_{l}'})}{|\boldsymbol{r-r_{l}'}|}+\frac{H(\boldsymbol{r_{l+1}'})}{|\boldsymbol{r-r_{l+1}'}|}\Big]d_{l,l+1}\label{eq8}
\end{equation}
where $L'(t)$ is uniformly sampled ($M$ points) across the angular sector covering the circular region-of-interest (ROI) of radius $R_{i}$ (Fig.~\ref{Fig1}(b)).
Assuming uniform SoS distribution of the medium, $L'(t)$ in Eq.~(\ref{eq6}) is an arc of a circle centered on the position of transducer element.
Thus, the segment $d_{l,l+1}$ of curve $L'(t)$ can be expressed as: 
\begin{equation}
    d_{l,l+1} = \left\{ \begin{aligned}
        &\frac{\alpha}{M-1}ct & 1 \leq l \leq M-1 \\
        &  0 & l=0, l=M\\
    \end{aligned}
    \right. \label{eq9}
\end{equation}
where $l$ indexes the segments, and $\alpha$ represents the opening angle of the sector covering the whole ROI viewing from transducer, calculated as:
\begin{equation}
    \alpha = 2 arcsin(\frac{R_{i}}{R_{t}})\label{eq10}
\end{equation}
in which $R_{t}$ is the radius of ring-shape transducer. 
Eq.~(\ref{eq8}) can be reduced to
\begin{equation}
    I(t) \approx  \frac{1}{2}\sum _{l=1} ^{M} \Big [\frac{H(\boldsymbol{r_{l}'})}{|\boldsymbol{r-r_{l}'}|}\Big](d_{l-1,l}+d_{l,l+1})\label{eq11}
\end{equation}
Combining Eq.~(\ref{eq6}) and Eq.~(\ref{eq11}), the forward model (Eq.~\ref{eq3}) can be formulated in matrix form as:
\begin{equation}
    \textbf{p} = \textbf{A}\textbf{H}\label{eq12}
\end{equation}
where $\textbf{p}$ is the model predicted pressure signals collected by the transducers, $\textbf{A}$ is the measurement matrix, and $\textbf{H}$ is the vectorized image to reconstruct.
With the MB method, PA images $\textbf{H}$ are then reconstructed by solving the inverse problem by minimizing the error between $\textbf{p}$ and experimentally measured signal $\textbf{p}_{m}$, i.e.
\begin{equation}
    \textbf{H}_{sol} =  arg \min_{\textbf{H}}||\textbf{p}_{m}-\textbf{A}\textbf{H}||_{2}^{2} + \lambda R(\textbf{H})\label{eq13}
\end{equation}
where $R(\textbf{H})$ and $\lambda$ are the regularization term and regularization parameter, respectively. 

\subsection{Neural representation method}
We use implicit neural representation (INR) for the PACT images, i.e., the image is represented as a continuous implicit function by a neural network $\mathcal{M}_{\Theta}$ parameterized with $\Theta$:
\begin{equation}
    \mathcal{M}_{\Theta}:(x, y) \xrightarrow{} \mathcal{H}\label{eq14}
\end{equation}
Image reconstruction with the NR method is then converted to finding the optimal $\Theta$ minimizing the loss function:
 \begin{equation}
\mathcal{L} = arg \min_{\Theta}||\mathcal{F}(\mathcal{H})-\textbf{p}_{m}||_{2}^{2}+\eta R(\mathcal{H})
\end{equation}
Here, $\mathcal{F}$(·) represents the forward operator from heat distribution $\mathcal{H}$ to PA signals (Eq.~\ref{eq12}),  $\textbf{p}_{m}$ is the measured PA signals by the ring array, and $\eta$ is a hyper-parameter for the regularization term.
Thus, the network is trained in a self-supervised manner.


\subsubsection{Coordinates Selection}
To train the neural network, the PA signal must be related to the spatial coordinate of the PA source.
At a specific sampling moment $t_i$, the PA signal amplitude received by the transducer element at $(x_{0},y_{0})$ is contributed by the point sources locating on the curve $L'(t_i)$ (Fig.\ref{Fig1}b).
After discretization, their coordinate $(x,y)$ can be calculated as:  
\begin{equation}
\left\{
\begin{aligned}
x &= ct_{i}cos(\beta_{0}+j\frac{\alpha}{M-1})+x_{0}  \\
y &= ct_{i}sin(\beta_{0}+j\frac{\alpha}{M-1})+y_{0} 
\end{aligned}
\right.
\end{equation}
where $j \in [0,1, 2, ...M-1]$ indexes the points located on $L'(t_i)$, with $\beta_{0}$ defined as  
\begin{equation}
\beta_{0} = arctan(\frac{y_{0}}{x_{0}})
\end{equation}

\subsubsection{Position Encoding and Network Structure}
After determining the coordinates of the point sources, the coordinates are hash encoded by multi-resolution mapping\cite{muller2022instant}, and the encoded hash vectors are fed to a MLP, which has two hidden-layers with 128 neurons (Fig.~\ref{Fig1}(c)).
ReLu is selected as the activation function of the neurons and the output activation function is the Sigmoid function to normalize the output value. 
\subsubsection{Training}
During the training process, the initial learning rate was set to 0.001 and decreased for every 20 epochs with a momentum of 0.5, and Adam optimizer \cite{kingma2014adam} was used to minimize the loss function.
The network was implemented with the tiny-cuda-nn \cite{muller2022instant} framework in Python.

\subsection{Experiments}
\subsubsection{In silico}
A vessel-like structure sized $ 2.5\times2.5$ cm$^{2}$ was placed at the center of a ring-shaped transducer array (40 mm radius, 256 elements) in a medium of uniform SoS (1500 $m\cdot s^{-1}$).
The sampling frequency was set at 20 MHz and the size of the computational grids was $512 \times 512$ with a pixel size of 0.05 mm. 
Given the the initial thermal distribution of the vessel-like structure (Fig.~\ref{Fig3}e) and the aforementioned settings, the PA signals were calculated by the forward model (Eq.~\ref{eq12}) stated in Sec. \uppercase\expandafter{\romannumeral2} .

\subsubsection{In vitro}
Five agar phantoms (1$\%$ agar w/v, 6 cm diameter) of different embedding materials forming various shapes were prepared for experimental validation of the proposed method.
Phantom 1  (1$\%$ agar w/v) had three embedded black plastic spheres ($3$ mm diameter) to mimick the scenario of imaging the cross-section of blood vessels in the body.
We also embedded black tungsten wires (0.1 mm diameter) forming leaf branch (Phantom 2), delta (Phantom 3), heart (Phantom 4) and star (Phantom 5) shapes, to mimick the scenario of imaging the longitudinal view of blood vessels in the body.

\begin{figure}
\centerline{\includegraphics[width=\columnwidth]{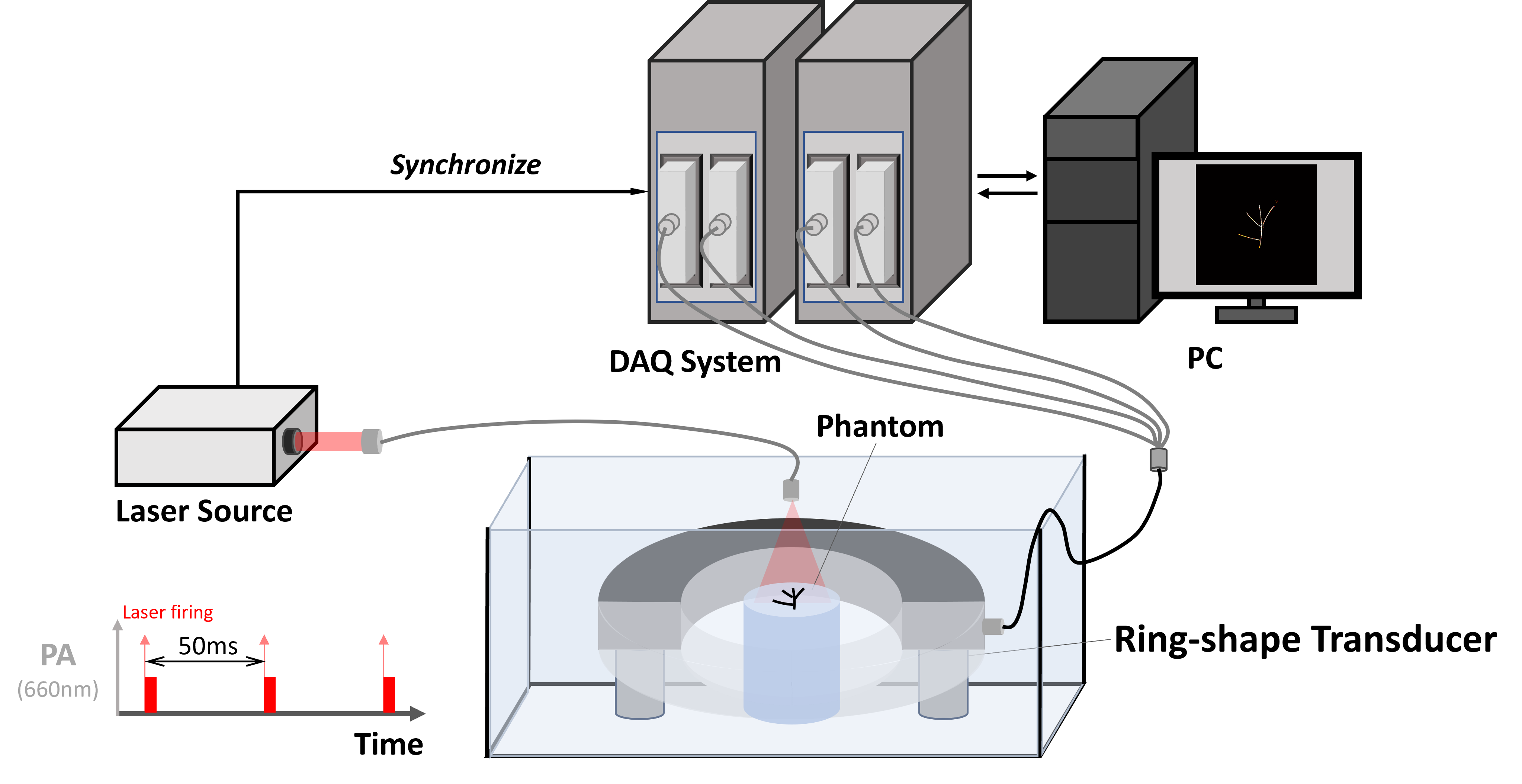}}
\caption{Phantom experiment setup. An optical fiber illuminates from top-view of the phantom and light spot covers the whole ROI. The laser source fires 20 times per second and delivers laser with 660 nm wavelength, and triggers vantage systems to acquire data synchronously at the same time.}
\label{Fig2}
\end{figure}

A 512-element ring transducer array of 80 mm diameter (center frequency: 5 MHz, Guangzhou Doppler Eletronic Technologies Co., Ltd, Guangzhou, China) was mated with two ultrasound research systems (Vantage 256, Verasonics, Kirkland, USA) for receiving PA signals in the experiments. 
The PA signals were initiated by a laser source (PHOCUS MOBILE, OPOTek Inc. USA)  (660 nm wavelength) with 20 Hz repetition rate, synchronizing the data acquisition with 20 MSPS sampling rate (resulting 1024 time samples).
For all the experiments, the light was emitted from the top-view of the imaging phantom in the de-ionized and degassed water, and the light spot covered the whole ROI (Fig.\ref{Fig2}). The SoS was obtained from the SoS-temperature relationship in pure water\cite{lubbers1998simple}.
During the experiments, the embedded material of the all the phantoms were place in the imaging plane of the ring array.
   
\subsection{Image reconstruction and performance evaluation}
For all the methods, the image reconstruction area was set as $2.5cm \times 2.5cm$ consisting of $512 \times 512$ pixels.
The signals consisting of 32, 64, 128, 256 and 512 projections were used to reconstruct image by different methods, respectively. 
For UBP method, the meaningless negative values on the image were set to zero for making a fair comparison. 
MB method was implemented following \cite{longo2022disentangling} using non-negative least square (NNLS) optimizing method with and the iteration number for solving the inverse problem was set to 50.
For the NR method, network training was stopped when the loss function is less than 0.0001.
Total Variation (TV) was used in the loss function of both the MB and NR methods, in which the regularization parameters $\lambda$ and $\eta$ were selected between 0 and 0.2 after trials.
All the above algorithms were implemented in Matlab 2020a on a workstation (Intel core i7, 14 cores at 2.90 GHz and 16GB memory) interfaced with a graphical processing unit (GPU, Nvidia Geforce RTX 4090 Ti).

For the simulation data, we utilized structural similarity index (SSIM)\cite{wang2004image} and peak signal-to-noise ratio (PSNR) as the metrics to evaluate the performance of different methods.
For the experimental data, we used signal-to-noise ratio (SNR) and contrast-to-noise ratio (CNR) as the evaluating metrics.

SSIM is defined as:
\begin{equation}
SSIM(f, gt) = \frac{(2\mu_{f}\mu_{gt}+C_{1})(2\sigma_{cov}+C_{2})}{(\mu_{f}^{2}+\mu_{gt}^{2}+C_{1})(\sigma_{f}^{2}+\sigma_{gt}^{2}+C_{2})}
\end{equation}
in which $f$ and $gt$ represent the reconstruction image and ground truth image, respectively. $\mu_{f}$ ($\sigma_{f}$) and $\mu_{gt}$ ($\sigma_{gt}$) are the mean (standard derivation) of reconstruction image $f$ and ground truth image, respectively, and $\sigma_{cov}$ is cross-covariance of $f$ and $gt$. The default parameter values of $C_{1}$ is 0.01, and $C_{2}$ is 0.03 \cite{b16}. The dynamic range of two parameters is 1. 

PSNR is defined as:
\begin{equation}
PSNR(f, gt) = 10log_{10}(\frac{I_{max}^{2}}{MSE})
\end{equation}
where $I_{max}$ represents the max value of $f$ and $gt$, and MSE is calculated by
\begin{equation}
MSE = \frac{1}{n^{2}}||f-gt||_{F}^{2}
\end{equation}
in which $||\textbf{·}||_{F}$ is Frobenius norm, and $n$ is the size of image.

SNR (unit in dB) is defined as:
\begin{equation}
SNR = 20log_{10}(\frac{\overline{I}_{signal}}{\sigma_{background}})
\end{equation}
where $\overline{I}_{signal}$ and $\overline{I}_{background}$ are the average amplitude of the selected imaging object region (green dash rectangles in Fig.\ref{Fig4}\&\ref{Fig5}) and background region (yellow dash rectangles in Fig.\ref{Fig4}\&\ref{Fig5}), respectively, $\sigma_{signal}$ and $\sigma_{background}$ represent the standard deviations of selected imaging object region and background region, respectively. 

CNR (unit in dB) is defined:
\begin{equation}
CNR = 20log_{10}(\frac{abs(\overline{I}_{signal}-\overline{I}_{background})}{\sigma_{background}})
\end{equation}
We computed the evaluation metrics with the images reconstructed by different methods from every projection number. 

\section{RESULTS}
\begin{figure*}
\centerline{\includegraphics[width=\textwidth]{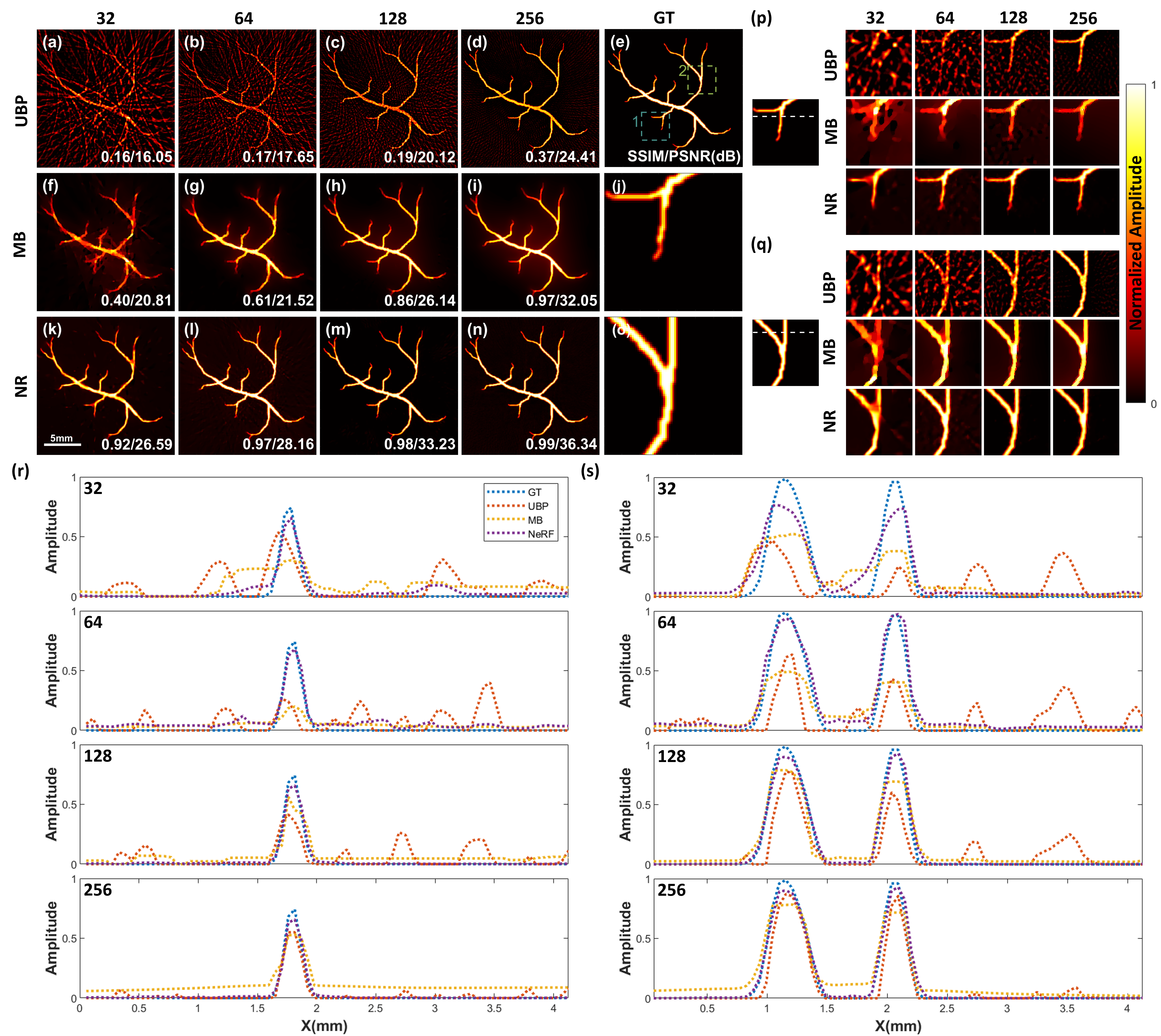}}
\caption{Results of simulation experiment. (a)-(d) Results of images reconstructed by UBP method from 32, 64, 128 and 256 projections respectively. (e), (j) and (o)  Ground truth. (f)-(i) Reconstructed results by MB method, the numbers of projection are the same as UBP method. (k)-(n) Reconstructed images by neural representation from different projection number respectively. (p) and (q) Zoom in specific regions of images. (r) and (s) the profile lines of images reconstructed by different methods from different projection number. NR, neural representation. }
\label{Fig3}
\end{figure*}

\subsection{Simulation Results}
For the vessel-like structure, the reconstructed images by the UBP, MB and NR methods using 32, 64, 128 and 256 projections are compared in Fig.\ref{Fig3}.
The parameters for the TV regularization term in the loss functions for MB and NR methods were selected as 0.01 and 0.02, respectively.
For 32, 64, 128 and 256 projections, the network converged after 100, 60, 40, and 20 training epochs, respectively.
The streak-type artifacts are noticeable in the images reconstructed by the UBP method, which became less visible as the projection number increased from 32 to 256 (SSIM/PSNR increased from 0.16/16.05 dB to 0.37/24.41 dB) (Fig.\ref{Fig3}(a-e)).

Compared to UBP, MB and NR methods remarkably removed the artifacts and improved the image quality.
As the the projection number increased from 32 to 256, the SSIM/PSNR of MB vs NeRF increased from 0.44/20.81 dB vs 0.92/26.59 dB to 0.97/32.05 dB vs 0.99/36.34 dB (Fig.\ref{Fig3}(f-i, k-o)).
A detailed comparison of two branches in the vessel-like structure (Fig.\ref{Fig3} e, p, q) and their line-profiles (Fig.\ref{Fig3} (r)-(s)) clearly demonstrated that NR had the best match with the ground-truth data.

\subsection{Phantom Experiments}
For phantom experiments, the images were reconstructed using 64, 128, 256 and 512 projections by different methods, respectively.
For MB method, $\lambda$ was set as 0.05, and the iteration number was set as 50. For NR method, The signal amplitude was normalized for training the network, and $\eta$ was set as 0.02. The network was convergent after 50 training epochs. 

\subsubsection{Plastic Sphere}
\begin{figure*}
\centerline{\includegraphics[width=\textwidth]{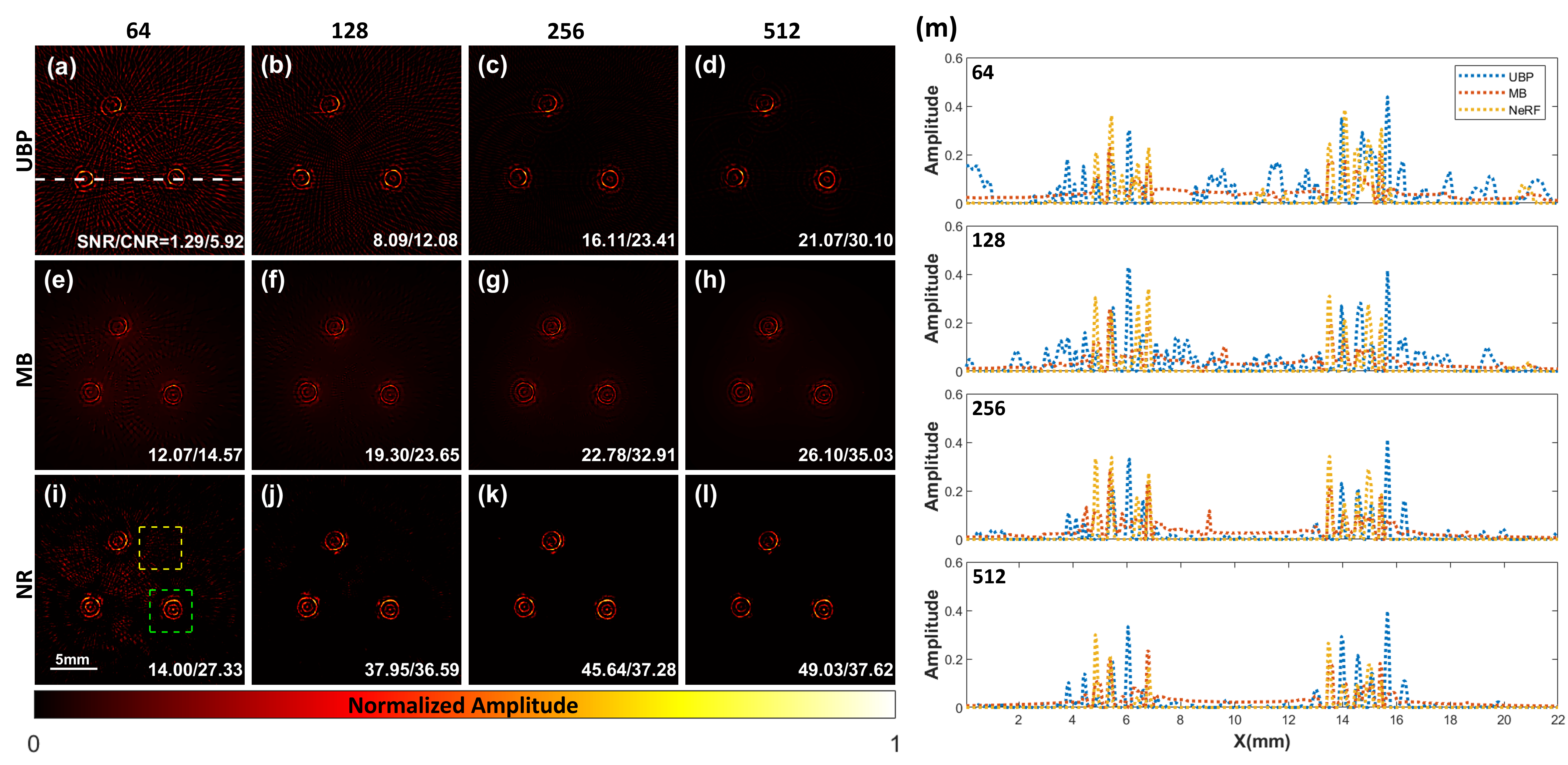}}
\caption{Results of phantom 1 experiment. (a)-(d) Results of images reconstructed by UBP method from 64, 128, 256 and 512 projections respectively. White dash line in (a) was selected to show the profile of pixels for each image. (e)-(h) Reconstructed results by MB method, the numbers of projection are the same as UBP method. (i)-(l) Reconstructed images by neural representation method from different projection number respectively. Green and yellow dash rectangles in (i) represent the signal region and background region which are selected to calculate CNR and SNR.  (m) The profile line of images reconstructed by different method from all projections. The CNR/SNR of image is set at the lower right corner of each image. NR, neural representation}
\label{Fig4}
\end{figure*}

In phantom 1, multiple rings were observed for each sphere in the images reconstructed by UBP, MB and NeRF (Fig.~\ref{Fig4}).
Consistent with previous observations in the simulated data, image quality of the images reconstructed by the MB and NeRF were better than that by the UBP method for the same projection number and adding more projection data for image reconstruction also improved the image quality for all different methods (Fig.\ref{Fig4}).
Specifically, the SNR/CNR of the images reconstructed by the UBP, MB and NeRF were 1.29 dB/5.92 dB, 12.07 dB/14.57 dB and 14.00 dB/27.33 dB (for 64 projections) and were consistently improved to 21.07 dB/30.10 dB, 26.10 dB/35.03 dB and 49.03 dB/37.62 dB, as the projection number increased to 512 projections.
Noticeably,in the images reconstructed by MB and NeRF, the two plastic spheres at the lower part in the images each had a dot recovered at the center of the rings, which was not observed in the images by UBP (Fig.\ref{Fig4}(a-l)).
When comparing the line profiles crossing the two spheres, it was observed that the profile lines in the UBP images had more fluctuations in the background area, while the MB and NR methods had more flat profile lines (Fig.\ref{Fig4}(m)).
Compare to the MB images, the profile lines in the NR images were closer to zero intensity in the background area between the two spheres.

\subsubsection{Tungsten Wires}
\begin{figure*}
\centerline{\includegraphics[width=\textwidth]{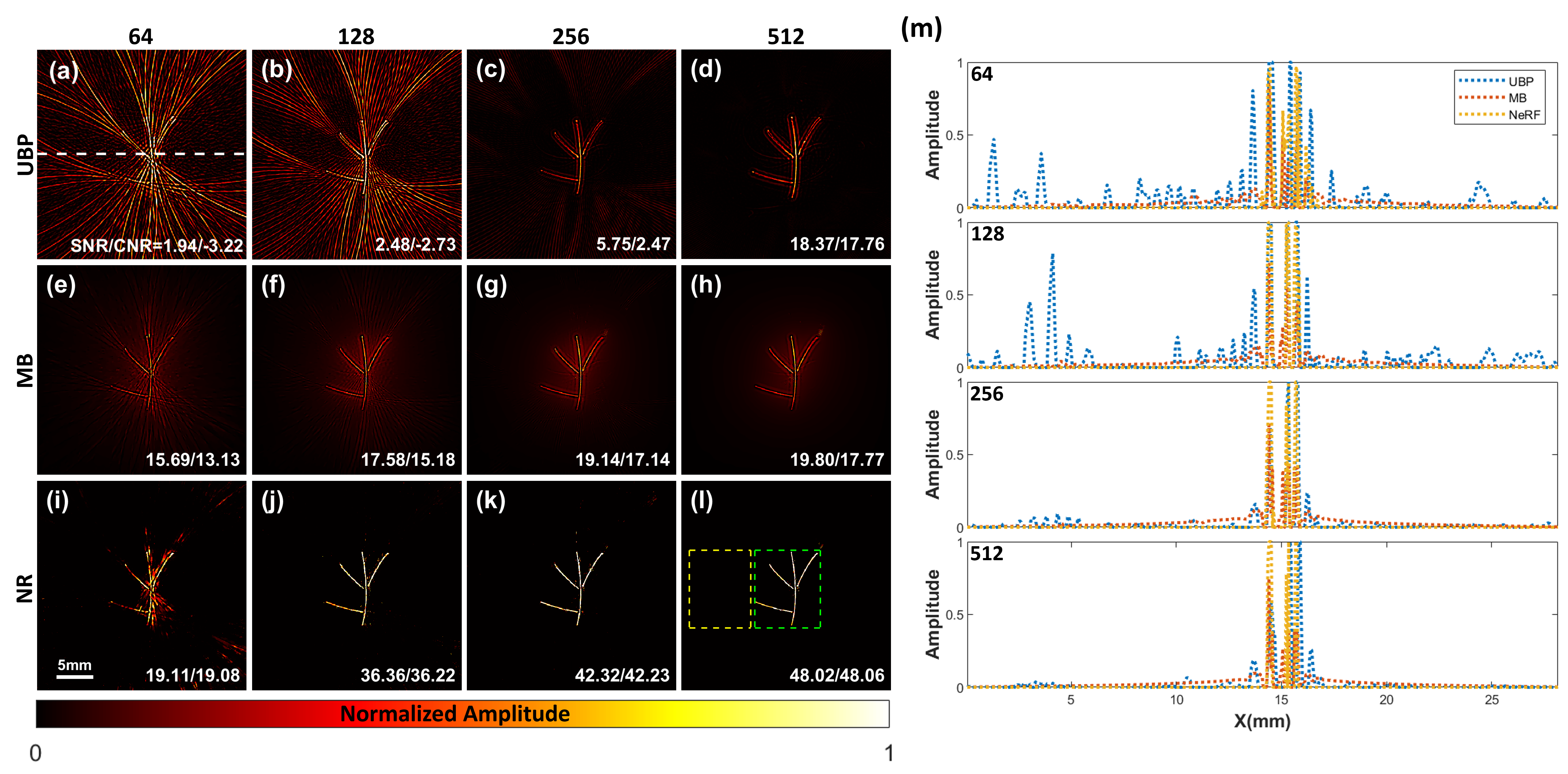}}
\caption{Results of phantom 2 experiment. (a)-(d) Results of images reconstructed by UBP method from 64, 128, 256 and 512 projections respectively. White dash line in (a) is the location of profile line. (e)-(h) Reconstructed results by MB method, the numbers of projection are the same as UBP method. (i)-(l) Reconstructed images by neural representation from different projection number respectively. Green and yellow dash rectangles in (l) represent the signal region and background region which are selected to calculate CNR and SNR. (m) The profile line of images reconstructed by different method from 64 and 128 projections.  The CNR/SNR of image is set at the lower right corner of each image. NR, neural representation}
\label{Fig5}
\end{figure*}

Consistent with the observation in phantom 1, in phantom 2 image reconstruction quality of the MB and NR methods were better than that by UBP method for the same projection number. 
For all the methods, the reconstructed image quality was improved with adding more projection data(Fig.~\ref{Fig5}).
Severe streak-type artifacts in the images reconstructed by UBP method when PA signals were very sparsely acquired (64 and 128 projections) (Fig.~\ref{Fig4}(a-b)).
These streak-type artifacts were largely mitigated by the MB and NR methods for the same projection number (Fig.~\ref{Fig4}(e-f, i-j)). 
For 256 and 512 projections, the streak-type artifacts were not obviously observed.
Comparing the line profiles along the middle of the object, it was observed that the profile lines of MB and NR methods had less fluctuations than that of UBP for all projection number (Fig.~\ref{Fig5}(m)).
Halo artifacts were consistently observed in the middle of the images reconstructed by the MB method, while this artifacts was removed in that of the NR method.
Thus, in the background area of images, the profile lines in the NR images were closer to zero intensity than that in the MB images.

Quantitatively, the SNR(CNR) of the images reconstructed by the UBP, MB and NR were 1.94 dB(-3.22 dB), 15.69 dB(13.13 dB) and 19.11 dB(19.08 dB) (for 64 projections) and were consistently improved as the projection number increased to 18.37 dB(17.76 dB), 19.80 dB(17.77 dB) and 48.02 dB(48.06 dB) (for 512 projections).
For the same number of projections, SNR(CNR) in the images reconstructed by the NR was consistently better than that of the UBP amd MB method, with 17.2-29.6 dB (22.3-29.8 dB) improvement over UBP and 3.4-28.2 dB (6.0-30.3 dB) improvements over MB for 64-512 projections.

To further demonstrate the performance of NR for sparse data, we compared the images reconstructed by UBP, MB and NR for Phantom 3-5 (Delta, Heart, Star) with 128 projections data(Fig.~\ref{Fig6}). For the imaging object with a radius of $1cm$, as observed, image reconstructed by UBP method remained streak-type artifacts in the background. With TV regularization term, MB method reconstructed images with less artifacts than UBP method. But the SNR and CNR of images were affected by the ill-posed problem. NR method removed most of the artifacts, and reconstructed the most distinct object structure.

Quantitatively, the SNR(CNR) of the images reconstructed by the UBP were 9.04 dB(8.36 dB), 10.57 dB(9.43 dB) and 7.36 dB(6.41 dB) for phantoms in different shape respectively. For the MB method, the SNR(CNR) of the images were 16.04 dB(14.21 dB), 17.40 dB(17.01 dB) and 15.01 dB(14.76 dB) for the same phantoms. And the NR method improved SNR(CNR) to 28.64 dB(27.59 dB), 29.01 dB(28.76 dB) and 26.28 dB(25.03 dB) respectively.
 \begin{figure}
\centerline{\includegraphics[width=\columnwidth]{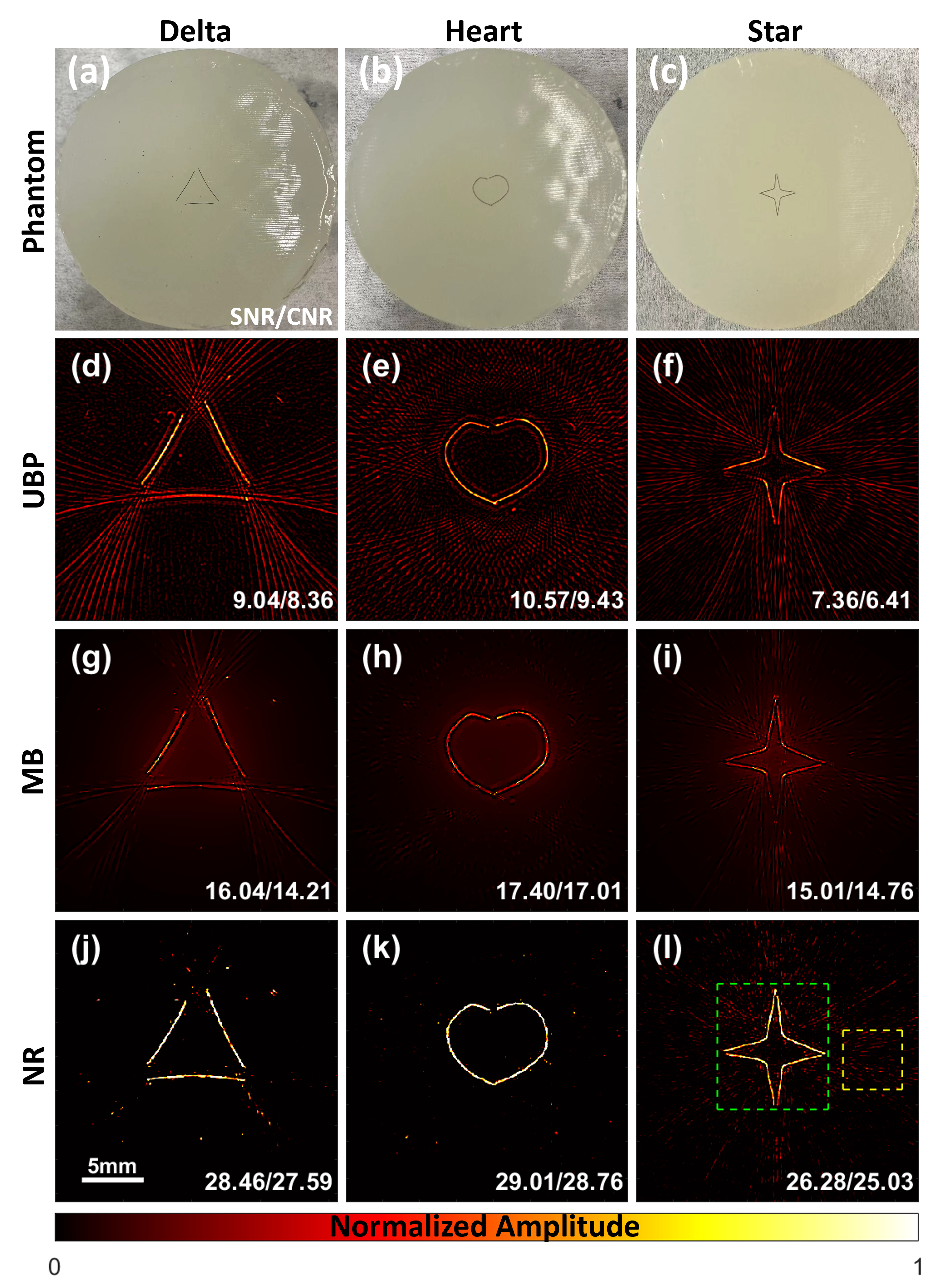}}
\caption{Experiment results for phantom 3, phantom 4 and phantom 5. (a)-(c)Phantom in different shapes. (d)-(f) Reconstruction images by UBP method from 128 projections. (g)-(i) Reconstruction images by MB method from 128 projections. (j)-(l) Reconstruction images by neural represantation method from 128 projections. Green and yellow dash rectangles in (l) represent the signal region and background region which are selected to calculate CNR and SNR. NR, neural representation}
\end{figure}
\label{Fig6}

\section{Discussion}
PACT combines the imaging depth of ultrasound image and contrast of medical optical image and has a great potential of being widely applied clinically \cite{xu2006photoacoustic}. However, existing reconstruction methods for PACT suffer a low image quality under the condition of sparse-view sampling. Developing imaging reconstruction method with high image quality and low imaging cost for PACT is therefore significant for the application of PACT in clinical practice. In this work, we proposed a neural representation framework to overcome the disadvantages of conventional PACT reconstruction methods under the circumstance of sparse PA acquiring condition, and reconstructs image from raw PA signals directly by learning the continuous representation of PACT image. We validated our method on simulation and experimental data with various sampling projections. The results show that the proposed method can effectively suppress artifacts under the condition of sparse sampling, and outperform UBP and MB methods \cite{b10,b12} under the same sampling condition. These results demonstrate the ability of our method for reducing the hardware cost and improving the image quality of PACT system.

Essentially, the image reconstruction for PACT is an inverse problem. Analytical method, such as UBP\cite{b10}, deduces an analytical solution of inverse problem from the forward propagation of ultrasound wave directly. UBP method is conventional for the ease of implementation and low time complexity when the spatial sampling condition satisfy the Nyquist sampling theorem. However, for sparse-view sampling, the image quality of analytical method is impacted by streak-type artifact severely. Compared to the UBP method, out method expresses the reconstructed image as a 2-D continuous function, and optimizes a MLP to fit this function under the constraint of prior knowledge, which make a better performance for sparse-view sampling.
Building a matrix expression of certain PA forward model, MB method\cite{b12} reconstruct PACT image by finding a numerical solution which minimize the loss function between the measured PA signals and the theoretical ones. For MB method, the interpolation method only take adjacent few pixels into consideration when calculating model matrix, which impact the accuracy of numerical solution. By applying hash encoding, our method interpolated a sampling pixel value from all vertexes of multi-resoluiton grids\cite{muller2022instant}, which makes a better performance than MB method. With no need of calculating model matrix, our method has a better time complexity than MB method. Moreover, for MB method, the increase of unknowns yields ill-posed problem for high-resolution image reconstruction. Benefiting from the continuous representation of MLP, NR method can reconstruct high-resolution image and not be affected by ill-posed problem. Some learning-based methods, such as AS-Net and diffusion model\cite{guo2022net, song2023sparse}, apply a pre-trained network model to remove artifacts from sparse-view images, which depends on a mass of high-quality training dataset. Compared to these learning-based methods, our method obtains high-quality PA images from sparse RF signals directly with no need of any reconstructed image, which reduces the complexity of training model and the difficulty of building dataset. And the quality of reconstructed image is not affected by dataset.

Simulation and phantom experiments are designed to compare the performance difference between our method and traditional methods. For simulation data (Fig.~\ref{Fig3}), UBP method suffers streak-type artifacts severely for the sparse-view sampling, which impact the image quality critically. With adding the prior knowledge provided by TV regularization term, MB method makes a better performance than UBP method under the same sampling condition, and improves the SSIM and PSNR of images markedly. NR method make a better performance than both above methods under the same sampling condition. For experimental data (Fig.~\ref{Fig4} and Fig.~\ref{Fig5}), SNR and CNR are calculated as the metric to measure image quality, and profiles of images are shown to compare the details of images. Consistent with the observation in simulation, NR method suppresses the artifacts in images and recovers images with a higher quality. We further compared the methods performance for diverse shapes under the sparse-view condition (Fig.~\ref{Fig6}). For the phantom size with a diameter of $1cm$, spatial sampling interval is $0.25mm$ (128 transducer elements) and bigger than half-wavelength ($0.15mm$), which not satisfy Nyquist criterion \cite{Hu2020Spatio}, so the case can be seen as a sparse sampling condition. Under this condition, NR method reconstruct images with less artifacts and higher SNR than UBP and MB methods.

The superiority of NR method over the compared method can be attribute to (1) the application of PA forward model deduced from PA wave propagation process, which map the collected data into plane coordinates; (2) the synergy between the continuous representation of an image offered by MLP and the external regularization term (TV) in the loss function. Our simulation and experimental results indicate that NR method outperforms traditional PACT reconstruction methods, e.g., UBP and MB, highlighting the value of implicit representation in solving inverse problem for ring-shaped transducer PACT reconstruction, especially for the sparse sampling condition.

While we showed that the NR method has a better performance to sparse-view PACT image reconstruction, it is noteworthy that the results attained under the condition that the medium is homogeneous. We must point that this method currently cannot make an outstanding performance to in-vivo imaging because of the heterogeneous SOS distribution inside the tissue. Moreover, a trained MLP represents a specific heat distribution. For reconstructing different images, the MLP need to be trained under the supervision of collecting RF signal, which increase the time complexity. In the future work, we will further take the effect of heterogeneous medium into consideration and explore a more efficient scheme to decrease the time cost of training MLP.  

\section{Conclusion}

In this paper, a framework based on neural representation was proposed, which reconstructed artifact-free PACT image from sparse-view PA data. Combing with the physical mechanism of photoacoustic forward model, a MLP network is trained in self-supervised fashion.  We validated the feasibility and performance of our proposed method by comparing other method through simulation data. In $in-vitro$ phantom experiments, our method shows superior performance compared with existing reconstruction methods. The neural representation method is still limited by the accuracy of forward model and time complexity of training, which may be further improved by building the propagation model of heterogeneous acoustic medium and finding more efficient ways to representing the plane coordinates. In the future work, we will improve the efficiency of our method and further generalize neural representation method to three dimensions for 3-D PA imaging.

\bibliographystyle{IEEEtran}
\vspace{12pt}
\footnotesize
\bibliography{IEEEabrv,Ref}

\end{document}